\documentclass[12pt]{article}

\usepackage{times, xcolor, graphicx, graphics, textcomp, amsmath, footmisc, array}
\usepackage[outdir=./Figures/]{epstopdf}
\usepackage[hidelinks]{hyperref}
\newcolumntype{P}[1]{>{\centering\arraybackslash}p{#1}}
\newcounter{lastnote}

\date{}

\begin{document} 


\baselineskip18pt

\begin{center}
\LARGE{Characterising Ion-Irradiated FeCr: Hardness, Thermal Diffusivity and Lattice Strain}\\
\bigskip
\normalsize{Kay Song$^{1}$\footnote[1]{Corresponding author: kay.song@eng.ox.ac.uk}, Suchandrima Das$^{1}$, Abdallah Reza$^{1}$, \\
	Nicholas W. Phillips$^{1}$, Ruqing Xu$^{2}$, Hongbing Yu$^{1}$, \\
	Kenichiro Mizohata$^{3}$, David E. J. Armstrong$^{4}$\footnote[2]{ david.armstrong@materials.ox.ac.uk}, Felix Hofmann$^{1}$\footnote[3]{felix.hofmann@eng.ox.ac.uk} }\\
\bigskip
\small{$^{1}$ Department of Engineering Science, University of Oxford, Parks Road, Oxford, OX1 3PJ, UK} \\
\small{$^{2}$ Advanced Photon Source, Argonne National Laboratory, 9700 South Cass Avenue, Argonne, IL 60439, USA} \\
\small{$^{3}$ University of Helsinki, P.O. Box 64, 00560 Helsinki, Finland} \\
\small{$^{4}$Department of Materials, University of Oxford, Parks Road, Oxford OX1 3PH, UK}
\\

\end{center}

\begin{center}
\textbf{Graphical Abstract}
\begin{figure}[h!]
	\includegraphics[width=0.8\textwidth]{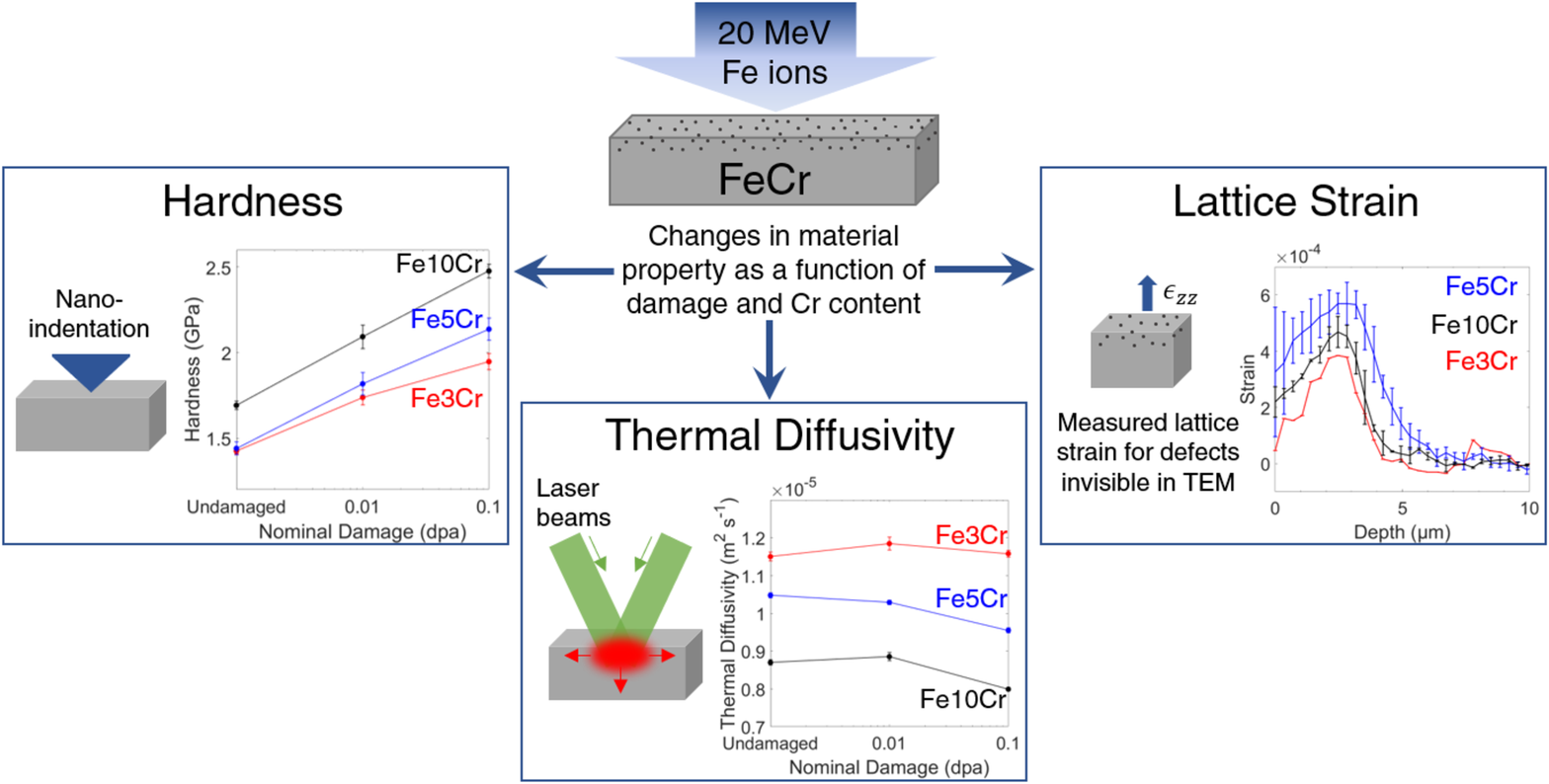}
	\centering
\end{figure}
\end{center}

\begin{abstract}
Ion-irradiated FeCr alloys are useful for understanding and predicting neutron-damage in the structural steels of future nuclear reactors. Previous studies have largely focused on the structure of irradiation-induced defects, probed by transmission electron microscopy (TEM), as well as changes in mechanical properties. Across these studies, a wide range of irradiation conditions has been employed on samples with different processing histories, which complicates the analysis of the relationship between defect structures and material properties. Furthermore, key properties, such as irradiation-induced changes in thermal transport and lattice strain, are little explored. 

Here we present a systematic study of Fe3Cr, Fe5Cr and Fe10Cr binary alloys implanted with 20 MeV Fe$^{3+}$ ions to nominal doses of 0.01 dpa and 0.1 dpa at room temperature. Nanoindentation, transient grating spectroscopy (TGS) and X-ray micro-beam Laue diffraction were used to study the changes in hardness, thermal diffusivity and strain in the material as a function of damage and Cr content. Our results suggest that Cr leads to an increased retention of irradiation-induced defects, causing substantial changes in hardness and lattice strain. However, thermal diffusivity varies little with increasing damage and instead degrades significantly with increasing Cr content in the material. We find significant lattice strains even in samples exposed to a nominal displacement damage of 0.01 dpa. The defect density predicted from the lattice strain measurements is significantly higher than that observed in previous TEM studies, suggesting that TEM may not fully capture the irradiation-induced defect population. 

\end{abstract}


\section*{Introduction}
In next-generation fusion and fission reactors, reduced activation ferritic/martensitic (RAFM) steels are likely to be used as the main structural material \cite{Baluc2004}. They are chosen for their high thermal conductivity and resistance to swelling compared to austenitic steels \cite{Ehrlich2001}. In service, the steel components will be exposed to temperatures up to 550$^{\circ}$C and intense irradiation by 14 MeV fusion neutrons \cite{Ehrlich1999}. For the optimisation of steel composition and operational safety, it is important to understand the defects formed and the microstructural changes due to irradiation, and the associated changes in thermomechanical properties.

To gain fundamental insight into the damage and defect population in steels due to irradiation, the study of iron-chromium (FeCr) binary alloys is useful, as it eliminates microstructural complications from other alloying elements. Ion implantation has become a widely used method to mimic neutron damage without the time and activation disadvantages of using neutrons in fission reactors \cite{Was2014, Harrison2019, Das2019}. Several past studies \cite{Xu2009, Jenkins2009, Prokhodtseva2013, Hardie2013} have shown the effective use of such ion-irradiated binary alloys with a range of Cr content to simulate the neutron irradiation conditions at various stages of component lifetime on differently composed steels.

There have been a number of transmission electron microscopy (TEM) investigations conducted on ion-irradiated FeCr alloys to directly image irradiation-induced defects. Ion-
implantation causes defects mainly in the form of dislocation loops, of both vacancy and interstitial types \cite{Jenkins2009, Yao2008, Hernandez-Mayoral2008}. Many studies have found that a `threshold dose' of around 0.01 dpa is required before visible defects appear \cite{Yao2008, Jenkins1978, Kirk1987}. This has been attributed to the need for an overlap of damage cascades before they form visible defect loops \cite{Jenkins2009, English1985}. However, a more recent study has found visible damage in TEM from damage as low as 0.0015 dpa \cite{Schaublin2017}. Higher damage results in the coalescence of loops as well as loop aggregation to form more complex microstructures such as loop strings \cite{Hernandez-Mayoral2008}. Cr has been observed to reduce loop mobility, leading to a higher number density of observable defects with smaller sizes compared to the case of pure Fe \cite{Prokhodtseva2013, Hernandez-Mayoral2008}. Irradiation temperature has also been observed to affect defect accumulation, with irradiation at 500$^{\circ}$C resulting in larger loops of only \textbf{b} = $\frac{1}{2}\langle 1 0 0 \rangle$ whereas 300$^{\circ}$C irradiation, at the same dose, produced smaller loops mostly with \textbf{b} = $\frac{1}{2}\langle 1 1 1 \rangle$ \cite{Jenkins2009}.

The effects of irradiation-induced defects on the mechanical properties of FeCr alloys have been studied with nanoindentation. Irradiation hardening has been observed in FeCr, and the amount of hardening increases with Cr content following irradiation at room temperature \cite{Heintze2011}. It has also been reported that the hardening effect saturates at damage above 2 dpa for Cr content greater than 5\% \cite{Hardie2013}. Another study found that hardness saturates at 1 dpa and above for Cr content $>$ 9\% \cite{Heintze2011}. Both of these studies were conducted with an irradiation temperature of 300$^{\circ}$C. It has also been found that hardening actually decreases with increasing dose rate, between $3\times10^{-5}$ to $6\times10^{-4}$ dpa/s, for FeCr alloys, which has been attributed to the clustering of Cr into precipitates \cite{Hardie2013a}. However, many of the existing TEM and nanoindentation studies on ion-irradiated FeCr were conducted on samples with different composition and processing history, as well as under different nominal irradiation conditions (total damage, dose rate and implantation temperature). As such, it is currently unclear how well these results can be compared and correlated with each other.

There are other important material properties that have not been extensively studied for ion-implanted FeCr. For example, it is crucial to study the effect of irradiation on the thermal diffusivity of FeCr as this will give a more accurate prediction of the temperatures expected for different reactor structural components and the amount of cooling required for safe operation. Little is currently known about the effects of irradiation on the thermal properties of Fe-based alloys, with one study finding negligible changes to the thermal conductivity of ASTM A533 grade B class 1 steel (ferritic) after neutron irradiation with 2.4 $\times 10^{23}$ n/m$^{2}$ ($\sim$ 0.036 dpa) \cite{Williams1983}. However, in tungsten, another BCC metal commonly studied under fusion-relevant conditions, self-ion irradiation to damage levels as low as 0.1 dpa has resulted in a 55\% degradation in thermal diffusivity \cite{Reza2020}. The degradation in thermal properties results from irradiation defects in the material lattice acting as electron scattering sites \cite{Ziman2001}. An interesting observation is that the defects observed with TEM do not account for all the measured changes to thermal diffusivity in tungsten, suggesting that small defects that TEM is not sufficiently sensitive to probe, have a significant effect on thermal and material properties following irradiation \cite{Zhou2006}. 

Irradiation-induced lattice swelling, and the associated stresses and strains, must also be considered for the design of structural reactor components \cite{Zinkle2000, Zinkle2013}. Previous experiments conducted on helium-implanted tungsten found large irradiation-induced strains and suggested higher defect retention per injected ion at lower doses \cite{Das2018a}. This shows that not only is lattice strain important to consider for reactor materials, but also that the effect of dose and damage history can significantly alter the damage response of a material. There has been many studies conducted on neutron and ion-irradiation induced void swelling in steels and FeCr \cite{Gelles1982, Gelles1995, Katoh1995, Sencer2000, Bhattacharya2016}. They have examined macroscopic swelling, often resulting in volumetric changes to the material, caused by the formation of voids and cavities, $>$10 nm in diameter,  after irradiation damage of tens to hundreds of dpa. This is different from the study of lattice swelling, which examines the microscopic strain induced by atomic defects \cite{Hofmann2015, Dudarev2018a}, and to date, little is known about the lattice swelling and strains in ion-irradiated FeCr. 

Here we address these open questions with a systematic study of FeCr with different Cr concentrations (3, 5 and 10 wt\% Cr) subjected to different irradiation damage levels, from 0.01 dpa to 0.1 dpa, at room temperature. Experimental characterisation of hardness, thermal diffusivity and lattice strain are conducted and the resultant trends discussed in terms of the underlying damage microstructure.

\section*{Methods and materials}
\subsection*{Materials and preparation}
High purity polycrystalline samples of FeCr containing 3, 5 and 10 wt\% Cr respectively were produced under the European Fusion Development Agreement (EFDA) programme (contract no. EFDA-06-1901) \cite{Coze2007, Fraczkeiwicz2011}. The alloys were prepared by induction melting under a pure argon atmosphere, then forging at 1150$^{\circ}$C. This was followed by a cold reduction of 70\% and heat treatment for 1 hour at 750$^{\circ}$C for Fe3Cr and Fe5Cr, and at 800$^{\circ}$C for Fe10Cr. The materials were then air cooled and delivered in the recrystallised state. The impurity content of the alloys is included in Appendix A. The alloy bars were sectioned with a diamond saw into samples $\sim 5\times5\times0.7$ mm$^{3}$ in size. 

The sample surfaces were mechanically ground with SiC paper, diamond suspension, then colloidal silica (0.04 $\mu$m). The final polishing step was electropolishing with 5\% perchloric acid in ethanol at -40$^{\circ}$C, using a voltage of 28 V and current of 0.3 A. The polishing time was on the order of 3 - 4 minutes, depending on the composition, until the surface deformation accumulated from the previous polishing steps had been removed. This was determined by visual assessment under optical microscopy. An unimplanted reference sample was retained for each Cr content, and ion implantation was performed for the remaining samples.

\subsection*{Ion implantation}
The samples were implanted with 20 MeV Fe$^{3+}$ ions at room temperature using a 5 MV tandem accelerator at the Helsinki Accelerator Laboratory. The beam size was approximately 5 mm and was rastered over a sample area of 10 $\times$ 10 mm$^{2}$ to achieve uniform implantation across the surface. The implantation chamber was held under vacuum at $8\times 10^{-7}$ mbar. 

\begin{figure}
\includegraphics[width=0.6\textwidth]{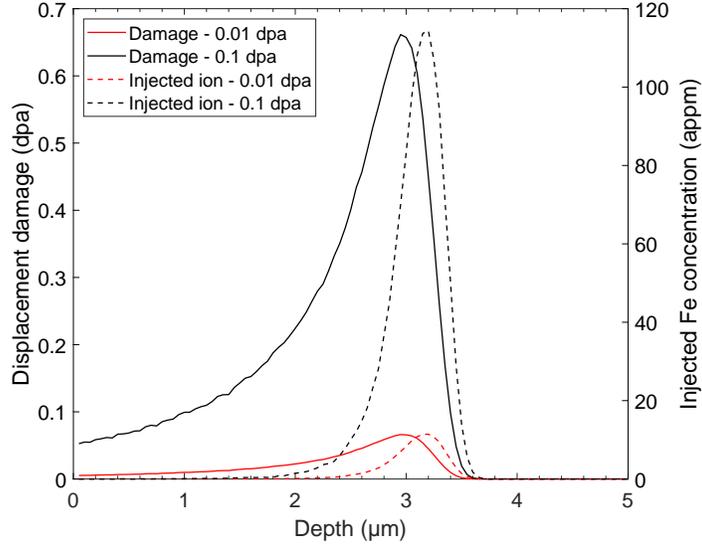}
\centering
\caption{Damage profiles (solid lines) predicted by SRIM for the two nominal damage levels of 0.01 dpa (red) and 0.1 dpa (black), which is the average damage in the first 2 $\mu$m below the surface. Also shown are the calculated injected ion concentrations (dashed lines) from the implantation.} 
\label{fig:SRIM}
\end{figure}

Figure \ref{fig:SRIM} shows the damage profile calculated using the Stopping and Range of Ions in Matter (SRIM) code \cite{Ziegler2010} using the Quick K-P calculation model with Fe ions and pure Fe as the target, with a displacement energy of 30 eV \cite{Olsson2016}. Cr has a similar displacement energy and density to Fe and molecular dynamics simulations have shown that there are no significant differences between the threshold displacement energies in FeCr and pure Fe \cite{Juslin2007}. Therefore the same damage profile is assumed for all samples of the same irradiation condition.

Two different fluences, $5.3 \times 10^{13}$ cm$^{-2}$ and $5.3 \times 10^{14}$ cm$^{-2}$ were used to achieve an average damage level of 0.01 dpa and 0.1 dpa, respectively, in the first 2 $\mu$m below the surface. This depth was chosen to calculate the nominal damage as it is before the sharp increase in damage level at the end of the implantation range (see Figure \ref{fig:SRIM}). These will be referred to as the `nominal damage level' from here onwards. The samples of different Cr composition were implanted at the same time for each nominal damage level at a dose rate of $2 \times 10^{-5}$ dpa s$^{-1}$.  We took advantage of the graduated damage profile when using depth-resolved techniques (discussed later in Micro-beam Laue Diffraction), allowing for a range of damage levels between 0 to 1 dpa to be probed. The damage level for a specific depth will be referred to as the `damage-at-depth'.


\subsection*{Nanoindentation}
Nanoindentation was performed on a MTS Nano Indenter XP with a Berkovich tip calibrated on fused silica with a known elastic modulus of 72 GPa. Continuous stiffness measurements (CSM) were made to a depth of 1 $\mu$m below the surface. 
Indents were made at a strain rate of 0.05 s$^{-1}$, a CSM frequency of 45 Hz and harmonic amplitude of 2 nm. At least 7 indents were performed on each sample, on grains within 10$^{\circ}$ of $\{$001$\}$ out-of-plane orientation, identified with EBSD before nanoindentation. Indents were spaced at least 50 $\mu$m from each other and grain boundaries.

\subsection*{Transient Grating Spectroscopy}
Transient Grating Spectroscopy (TGS) is able to provide rapid and non-destructive measurements of thermal diffusivity in thin surface layers. The technique is described in detail in \cite{Reza2020, Hofmann2015a, Dennett2017, Kading1995}. Two short laser pulses (pump beams of 0.5 ns at 532 nm) are directed at the surface of the sample at a fixed angle forming an interference pattern with a specific wavelength, $\lambda$, on the sample surface. At the positions of constructive interference, the energy of the lasers is absorbed by the sample and causes local thermal expansion, which creates a spatial displacement grating on the surface. As the heat diffuses from maxima to minima, and into the material bulk, the transient thermal grating decays. This response can be measured by diffracting a second laser beam (probe beam, continuous at 559.5 nm) off the `transient' grating on the sample surface. The decay of the diffracted signal intensity can be interpreted in terms of the thermal diffusivity of the material. The thickness of the probed layer is approximately $\frac{\lambda}{\pi}$ \cite{Kading1995}.

In this study, a grating wavelength of $\lambda = 5.707 \pm 0.001$ $\mu$m was used to obtain a probed depth of approximately 2 $\mu$m, in order to study the irradiated layer. The average pump beam power was 1.5 mW operating at 1 kHz, and the probe beam power was 22 mW, at 1 kHz with a 25\% duty cycle. The reflectivity of pure Fe for this wavelength is $\sim$50\% \cite{Bennett1981, Coblentz1910}. This corresponds to an approximate absorbed energy of 0.75 mW and 11 mW from the pump and the probe beams respectively. The pump beam and probe sizes were respectively 140 $\mu$m and 90 $\mu$m (1/$e^{2}$ width). 25 spot measurements were made on each sample across an area of 1 mm$^{2}$. At each location, 20000 laser pulses, and the corresponding signal traces, were recorded. The measurements were performed at room temperature under vacuum ($\sim 1\times 10^{-3}$ mbar).

\subsection*{Micro-beam Laue Diffraction}
Lattice swelling in the implanted layer of the FeCr samples was measured using micro-beam Laue diffraction at the 34-ID-E beamline, Advanced Photon Source, Argonne National Laboratory, USA. Using the Differential Aperture X-ray Microscopy (DAXM) technique, described in \cite{Das2018a, Larson2002, Hofmann2013}, depth-resolved lattice strain measurements were made. A thin platinum wire was scanned across the sample surface during the measurements. The diffraction patterns from subsequent positions of the wire were compared, with the wire edge positions acting as the `differential aperture', allowing for the resolution of signals from different depths. Measurements were made to allow reconstruction of signals from as deep as 15 $\mu$m below the surface along the beam's penetration direction (45$^{\circ}$ to the sample surface). The size of the beam on the sample surface was approximately $400 \times 200$ nm$^{2}$ and the depth resolution was estimated to be $\sim$1 $\mu$m.

On each sample, grains within 10$^{\circ}$ of $\{$001$\}$ out-of-plane orientation, previously identified with EBSD, were measured by diffraction of a monochromatic X-ray beam. At least 2 points were measured on each sample, except for Fe3Cr 0.1 dpa where only 1 point was measured as strong material texture restricted the number of grains with the desired orientation. A $\{$00n$\}$ reflection with energy between 12 - 18 keV was selected and scanned across an average photon energy range of $\sim$40 eV with a step size of 1 eV. Only the out-of-plane strain was measured as previous experiments on tungsten found that the in-plane strains from ion implantation are close to zero \cite{Das2018a, Hofmann2015}. Analysis of the diffraction patterns was performed with the LaueGo \cite{Tischler2020} to extract lattice strain as a function of depth in the sample.

\section*{Results and Discussion}
\subsection*{Irradiation Hardening}
\begin{figure}[h!]
	\includegraphics[width=\textwidth]{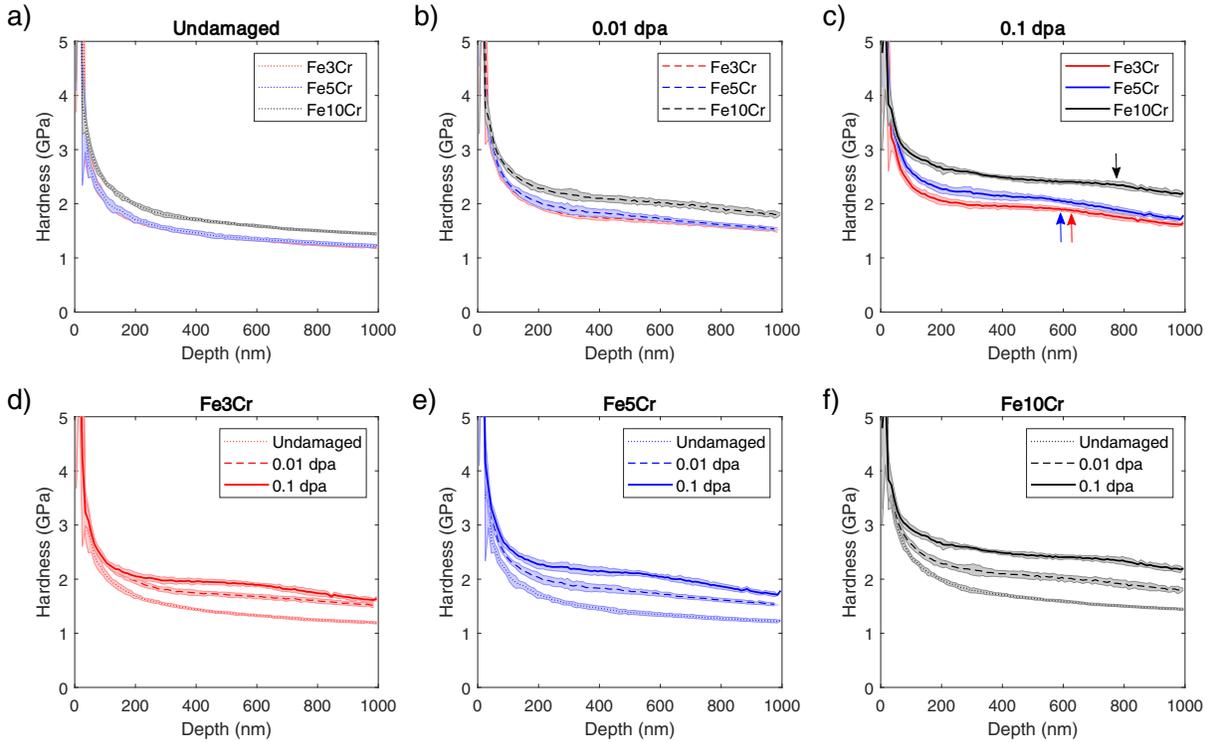}
	\centering
	\caption{Hardness of the samples as a function of indented depth from CSM nanoindentation measurements. a), b) and c) show data grouped by damage level. The arrows in c) indicate the depths at which the hardness curve of the corresponding Cr content (same colour) changes slope. d), e) and f) show data grouped by Cr content. The data shown consist of averaged hardness from all indents taken on each sample, with the error bars representing $\pm$ one standard deviation. Note that the damage layer as predicted by SRIM calculations extends to a depth of 3.5 $\mu$m.} 
	\label{fig:Hardness_Depth}
\end{figure}

From nanoindentation in CSM mode, hardness as a function of indentation depth was obtained for all the samples (Figure \ref{fig:Hardness_Depth}). The data shown is an average of all the measurements taken on each sample and the error bars represent $\pm$ one standard deviation from each set of measurements. The indentation size effect \cite{Nix1998}, resulting in high hardness values measured in the first 200 nm below the surface, is very clear in all the measurements. Data from the unimplanted samples is shown in Figure \ref{fig:Hardness_Depth}a. The hardness changes smoothly as a function of depth as expected for a homogenous bulk material. Fe10Cr shows the highest hardness, and the difference in hardness between the samples of different Cr content is similar at all depths. 

Figure \ref{fig:Hardness_Depth}b shows that even at a low nominal damage level of 0.01 dpa, there is more spread in the hardness data for all Cr compositions compared to the undamaged reference samples. Figure \ref{fig:Hardness_Depth}c compares the effect of Cr at 0.1 dpa and it can be seen that Fe5Cr has hardened substantially more than Fe3Cr, up to 0.3 GPa more at a depth of 300 nm, considering the hardness values of the corresponding unirradiated samples are the same. 

A change in the slope of the hardness plots as a function of indentation depth is observed after 0.1 dpa of nominal damage for all Cr compositions. This feature is visible in Fe10Cr at around 750 nm, and Fe5Cr and Fe3Cr at 550 nm and 570 nm respectively (see arrows in Figure \ref{fig:Hardness_Depth}c). The depth at which the slope of the curve changes ($h_{c}$) is thought to correspond to the indentation depth at which the plastic zone extends into the unimplanted bulk material, which is softer \cite{Xiao2018}. The greater $h_{c}$ is, the shallower the plastic zone ahead of the indenter tip is. The Johnson model of plastic zone size gives \cite{Johnson1970}:
\begin{equation} \label{eqn:pz}
\frac{c}{a_{s}} = \left( \frac{2E}{3\sigma_{ys}} \right)^{1/3}
\end{equation}

where $c$ is the radius of the plastic zone, $a_{s}$ is the radius of the indenter imprint, $E$ is the elastic modulus and $\sigma_{ys}$ is the yield stress. This model was originally developed for a spherical indenter but it has been demonstrated to be valid for a Berkovich indenter if we substitute the ratio $\frac{c}{a_{s}}$ with $\frac{z_{p}}{h}$ where $z_{p}$ is the depth of the plastic zone directly beneath the indenter at an indentation depth $h$ \cite{Mata2006}. 

For FeCr, the elastic modulus does not change significantly with Cr content for Cr content $<$ 20\% \cite{Speich1972, Masumoto1971} and this is also seen in our measurements in Appendix B. It has also been found that the yield stress of FeCr is proportional to indentation hardness \cite{Heintze2011}. The hardness measurements in Figure \ref{fig:Hardness_Depth}a indicate that yield stress increases with Cr content. Therefore the ratio $\frac{z_{p}}{h}$ is expected to decrease with Cr content, which agrees with the current observations. Furthermore, upon substituting the experimental values into Equation \ref{eqn:pz}, both sides of the equation agree to within a factor of 3 (raw data included in Appendix C). This gives further confidence in the use of the Johnson model to explain our experimental observations.

Figure \ref{fig:Hardness_Depth}d, shows quite clearly for Fe3Cr that, due to the change in the hardness slope at shallow depths (around 550 nm), the hardness values for 0.1 dpa of nominal damage are already approaching the 0.01 dpa values at 1000 nm. The Fe5Cr samples (Figure \ref{fig:Hardness_Depth}e) and Fe10Cr samples (Figure \ref{fig:Hardness_Depth}) both show greater difference in hardness at all depths between different levels of damage than Fe3Cr. The difference in hardness at the greatest indentation depth of 1 $\mu$m increases with Cr content.

\begin{figure}[h!]
	\includegraphics[width=0.6\textwidth]{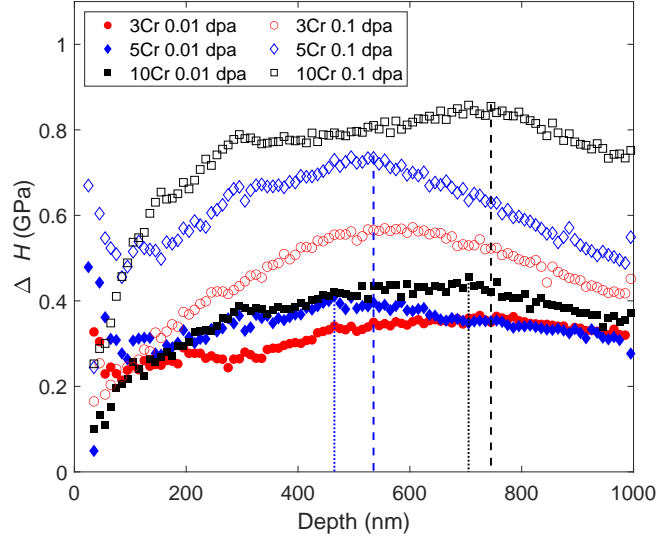}
	\centering
	\caption{The absolute change in hardness ($\Delta H$) compared to the undamaged samples of each composition. The dotted lines indicate the depth at which $\Delta H$ is maximum ($h_{c}$) for samples with 0.01 dpa of nominal damage (filled circles) and the dashed lines indicate $h_{c}$ for samples with 0.1 dpa of nominal damage (unfilled diamonds). } 
	\label{fig:Delta_Hardness}
\end{figure}

Figure \ref{fig:Delta_Hardness} shows the change in hardness ($\Delta H$), calculated as the difference between the irradiated samples and their corresponding undamaged counterpart of the same composition, as a function of indentation depth. For Fe5Cr and Fe10Cr, there is a clear peak in $\Delta H$ at $h_{c}$ for samples of both damage levels. For both Fe5Cr and Fe10Cr, $h_{c}$ is greater at 0.1 dpa of nominal damage than 0.01 dpa. Considering the 0.1 dpa samples are harder than 0.01 dpa, this suggests lower yield stress in 0.01 dpa samples \cite{Heintze2011}. Using the Johnson model \cite{Johnson1970}, lower yield stress would correspond to larger plastic zone in the indented material, which qualitatively agrees with the findings of lower $h_{c}$ in the 0.01 dpa samples compared to 0.1 dpa. 
The $\Delta H$ vs. indentation depth curves of Fe3Cr damaged to 0.01 dpa may exhibit a slightly larger $h_{c}$ value than the 0.1 dpa damaged sample, however, this is harder to discern compared to the case of Fe5Cr and Fe10Cr. As such, it has not been considered for analysis with the Johnson model.

\begin{figure}[h!]
\includegraphics[width=0.6\textwidth]{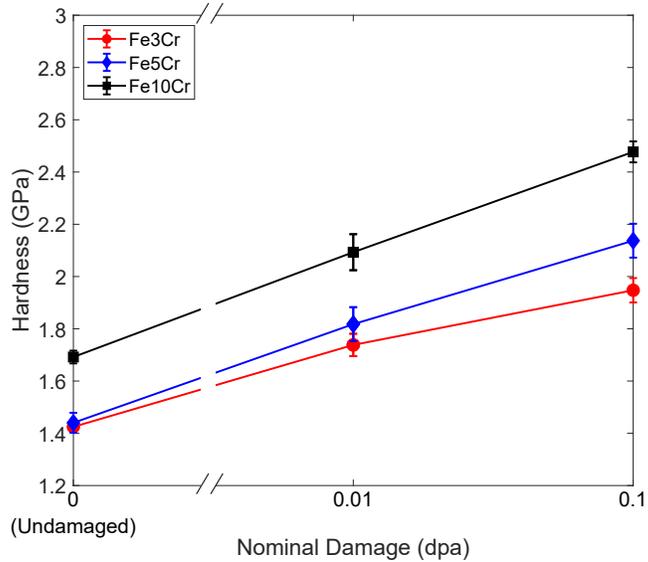}
\centering
\caption{Average hardness values for each sample averaged between indentation depths of 300 to 600 nm for Fe3Cr (red), Fe5Cr (blue) and Fe10Cr (black) at each damage level. Error bars represent $\pm$ one standard deviation from the measurements taken on each sample.} 
\label{fig:Hardness_Summary}
\end{figure}

Figure \ref{fig:Hardness_Summary} shows depth-averaged hardness values (300 to 600 nm) for all samples, with the error bars representing $\pm$ one standard deviation. This depth region was selected as it is deeper than the region dominated by the indentation size effect but below $h_{c}$, representing the hardness in the irradiated layer. All samples exhibit irradiation hardening following implantation and the hardness value at each damage level increases with Cr content. This is consistent with findings for up to 1 dpa of damage at room temperature for Fe2.5Cr, Fe9Cr and Fe12Cr \cite{Heintze2011}. 

It can be seen that the amount of hardening relative to the undamaged samples of the same composition is higher for Fe5Cr and Fe10Cr than for Fe3Cr at 0.1 dpa of nominal damage. The relative hardness increase for Fe3Cr, Fe5Cr and Fe10Cr is respectively 36\%, 48\% and 46\%. This agrees with previous nanoindentation measurements of 1 dpa damaged samples \cite{Heintze2011} and suggests a greater retention of defects at higher Cr content. This is also consistent with TEM findings of reduced defect mobility and increased number density of defects with increasing Cr content \cite{Prokhodtseva2013, Hernandez-Mayoral2008}.

We also note that the amount of hardening between 0.01 dpa and 0.1 dpa of nominal damage is greater for Fe5Cr and Fe10Cr than Fe3Cr.  This agrees with findings from previous studies where hardening increased with Cr content from 2.5\%Cr to 12.5\%Cr following room temperature irradiation \cite{Heintze2011}. Hardie \textit{et al.} \cite{Hardie2013} found that hardening increases slightly with Cr content at low dose ($\sim$2 dpa) but at a higher dose ($\sim$6 dpa), samples with Cr content $<$7\% exhibited much more irradiation hardening. However, these studies were performed with an irradiation temperature of 300$^{\circ}$C and at a higher damage level than those in the present study. As such, it would be of interest in follow up studies to conduct room temperature irradiation to a higher damage level in order to compare differences in hardening for different Cr content.

\subsection*{Thermal Diffusivity}
\begin{figure}[h!]
	\includegraphics[width=\textwidth]{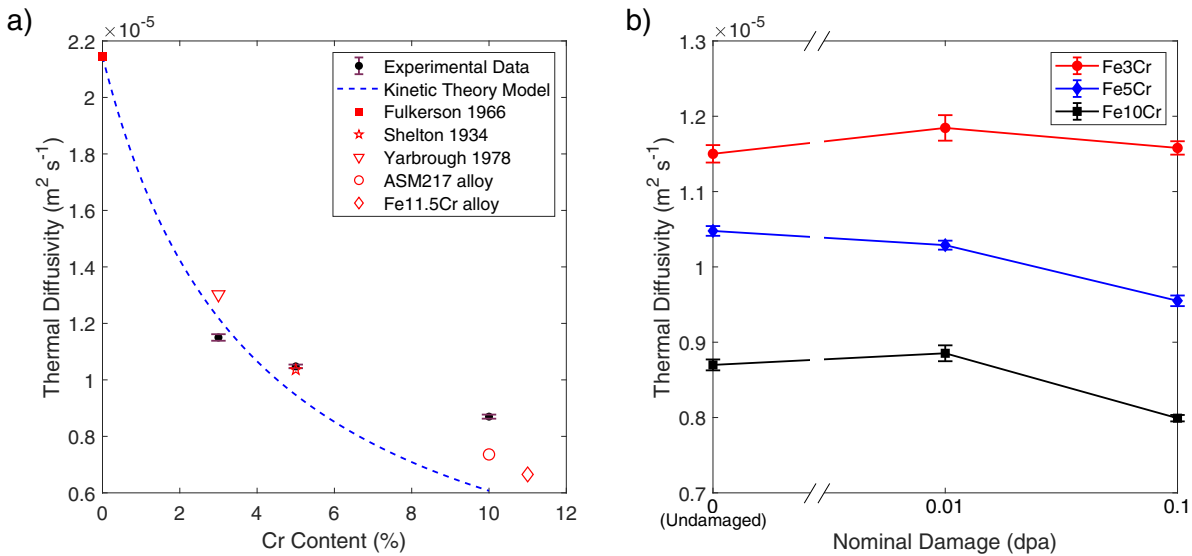}
	\centering
	\caption{a) The average thermal diffusivity values for undamaged samples (black circles). The predictions of thermal diffusivity based on the kinetic theory model (blue dashed line) and the experimental data from other studies on Fe and FeCr (red markers) are also shown. b) Average thermal diffusivity values for each sample with error bars representing $\pm$ one standard deviation from the 25 measurement points on each sample.} 
	\label{fig:TD_Summary}
\end{figure}
Figure \ref{fig:TD_Summary}a shows the thermal diffusivity, averaged over 25 measurement points, for the undamaged samples of different Cr content measured with TGS. The error bars represent $\pm$ one standard deviation computed from the measurement points. For Fe5Cr, the measured value agrees very well with the literature \cite{Sheiton1934}. For Fe3Cr, the agreement is good (within 10\%), a discrepancy that may be accounted for by some surface oxidation on our material \cite{Yarbrough1978} as low Cr content may be insufficient to prevent oxidation \cite{Pujilaksono2011}. For the thermal diffusivity of undamaged Fe10Cr, there is limited information in existing literature and comparisons were made with ASM217 alloy (8-10\% Cr, 0.2\% C and 1\% Si)\cite{ASMHandbookCommittee1990} and a Fe11.5Cr alloy (0.4\% C and 0.3\% Mn)\cite{Smithells1955}, giving 15\% and 25\% agreement, respectively. TGS values of thermal diffusivity have been validated for a number of other material systems \cite{Reza2020, Hofmann2015a, Alam1994, Xu2008}, which combined with the agreement with literature values for FeCr, gives confidence in our measurements.

A trend of decreasing thermal diffusivity with increasing Cr content is observed. This can be rationalised by considering Cr acting as `impurities' that scatter electrons, decreasing the electron scattering time $\tau_{e}$ \cite{Wang2019}. We can further analyse this by considering a kinetic theory model \cite{Hofmann2015a} that focuses on the contribution of electrons as the main heat carrier. Matthiessen's rule states that the total electron scattering rate $\sigma = \frac{1}{\tau_{e}}$ is the sum of the scattering rates from impurities (in this case Cr), phonons and other electrons, as long as the electron mean free path is larger than the separation between atoms \cite{Wiesmann1977, Mason2015}. Focusing on the contributions of Cr altering the scattering of electrons from the pure Fe case, the thermal diffusivity, $\alpha$, is \cite{Hofmann2015a}:
\begin{equation}\label{eqn: kinetic}
\alpha = \frac{C_{e}\nu_{f}{}^{2}}{3\rho c_{p}}\left( \frac{c_{Cr}}{\tau_{e,Cr}} + \frac{1-c_{Cr}}{\tau_{e,Fe}} \right)^{-1}
\end{equation}
where $C_{e}$ is the electronic heat capacity, $\nu_{f}$ is the Fermi velocity, $\rho$ is the density and $c_{p}$ is the specific heat capacity. $c_{Cr}$ is the atomic fraction of Cr, $\tau_{e,Fe}$ is the electron scattering time of pure Fe and $\tau_{e,Cr}$ is the scattering time of Cr `impurities', which is fitted from experimental data. $\tau_{e,Fe}$ is calculated from the thermal conductivity of pure Fe $\kappa_{Fe}$ \cite{Fulkerson1966}, given by \cite{Ashcroft1976}:
\begin{equation}
\kappa_{Fe} = \frac{1}{3} C_{e}\nu_{f}{}^{2}\tau_{e,Fe}
\end{equation}

The values of $C_{e}$, $\nu_{f}$, $\rho$, $c_{p}$ and $\kappa_{Fe}$ used for the fit are shown in Appendix D. The contribution of Cr to electron scattering can be fitted from our experimental data as $\tau_{e,Cr} = 0.35$ fs compared to $\tau_{e,Fe} = 92$ fs. 
This kinetic theory model gives a good fit to the experimental data at low Cr content, shown in Figure \ref{fig:TD_Summary}a. This is reasonable as the model assumes a dilute alloy. 

Figure \ref{fig:TD_Summary}b shows the thermal diffusivity of all the samples, of different Cr content and damage levels. For Fe3Cr, there is no significant change in the thermal diffusivity following ion implantation. However, for Fe5Cr and Fe10Cr, there is a significant decrease in thermal diffusivity ($\sim$ 8\%) after a nominal damage level of 0.1 dpa. This suggests a higher retention of defects in Fe5Cr and Fe10Cr, which qualitatively agrees with previous TEM observations of Cr reducing defect mobility \cite{Hernandez-Mayoral2008} and enhancing defect retention \cite{Prokhodtseva2013}.

At 0.01 dpa of nominal damage, both the Fe3Cr and Fe10Cr samples show a slight increase in thermal diffusivity compared to the respective undamaged samples of the same composition, though still within the error bars. This increase is surprising and the mechanism behind this is unclear. The only possible explanation is the clustering of Cr with damage but atom probe tomography measurements have shown that this only occurs at higher damage levels and irradiation temperature ($\sim$ 0.6 dpa at 300$^{\circ}$C) \cite{Hardie2013, Kuksenko2013}. It is worth noting that the error bars represent the standard deviation of the 25 spot measurements taken on each sample. Therefore, they are indicative of the variation in measurements on each sample rather than the absolute accuracy of the measurements themselves. 

TGS measurements on self-ion implanted tungsten have shown a 55\% decrease in thermal diffusivity following irradiation, and the effect saturates at 0.1 dpa of damage \cite{Reza2020}. It would be of interest to investigate this for FeCr by conducting TGS measurements on samples damaged to higher dpa. Given that the decrease in thermal diffusivity at 0.1 dpa of nominal damage is quite small, this is potentially good news for future reactors as the changes in thermal diffusivity due to defects may be too small, compared to the effect from the Cr content, to have an impact on predicted design purposes. However, the effect of alloying impurities on thermal diffusivity would need to be carefully considered in material design.


\subsection*{Lattice Swelling}
\begin{figure}[h!]
	\includegraphics[width=\textwidth]{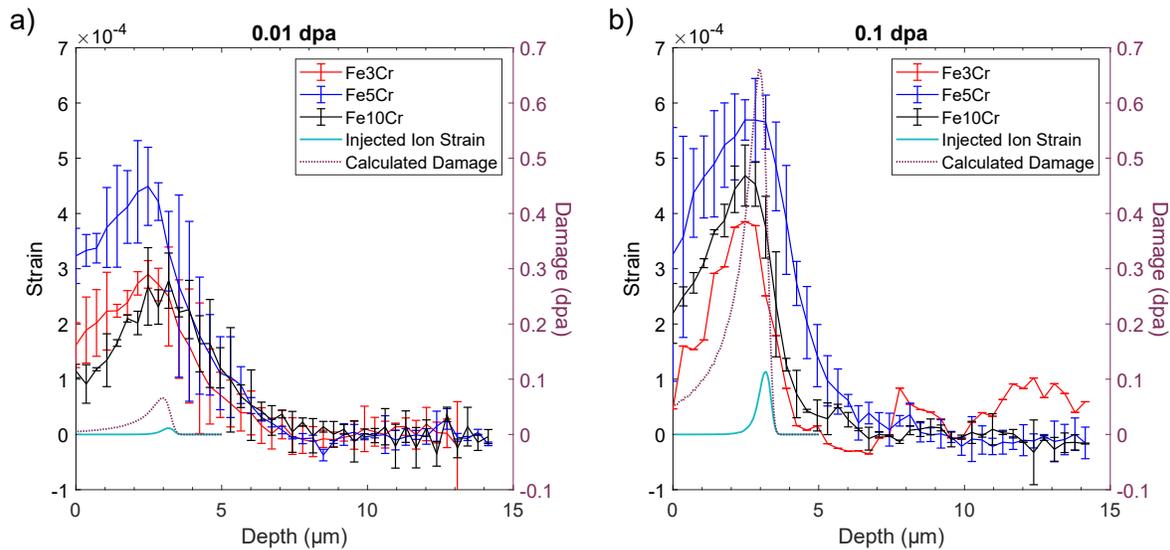}
	\centering
	\caption{Measurement of average strain as a function of depth perpendicular to the surface of each sample for each composition (Fe3Cr - red, Fe5Cr - dark blue and Fe10Cr - black) at a) 0.01 dpa of nominal damage and b) 0.1 dpa of nominal damage. The error bars represent $\pm$ one standard deviation, noting that only one measurement was taken for the Fe3Cr 0.1 dpa sample due to the lack of available grains with \{001\} out-of-plane orientation. Also shown in both plots is the predicted strain from injected ions if they all exist as $\langle 111 \rangle$ interstitials in the lattice (teal line) \cite{Hofmann2015, Ma2019} and the damage profile (dotted purple line), both obtained from SRIM calculations.} 
	\label{fig:Strain_Depth}
\end{figure}

Figure \ref{fig:Strain_Depth} shows the out-of-plane strains measured in the samples as a function of depth. The strains are greatest around $2.5 - 3$ $\mu$m depth, the region with the highest damage and injected ion concentration (as predicted by SRIM). The amount of strain rapidly decreases to zero at depths greater than 5 $\mu$m, suggesting that there is little diffusion of defects from the implanted layer into the bulk. It can also be seen that the strain contribution calculated from the injected ions of the implantation, assuming they all exist as $\langle 111 \rangle$ interstitial defects in the lattice \cite{Hofmann2015, Ma2019}, is small compared to the measured strain. 

\begin{figure}[h!]
\includegraphics[width=\textwidth]{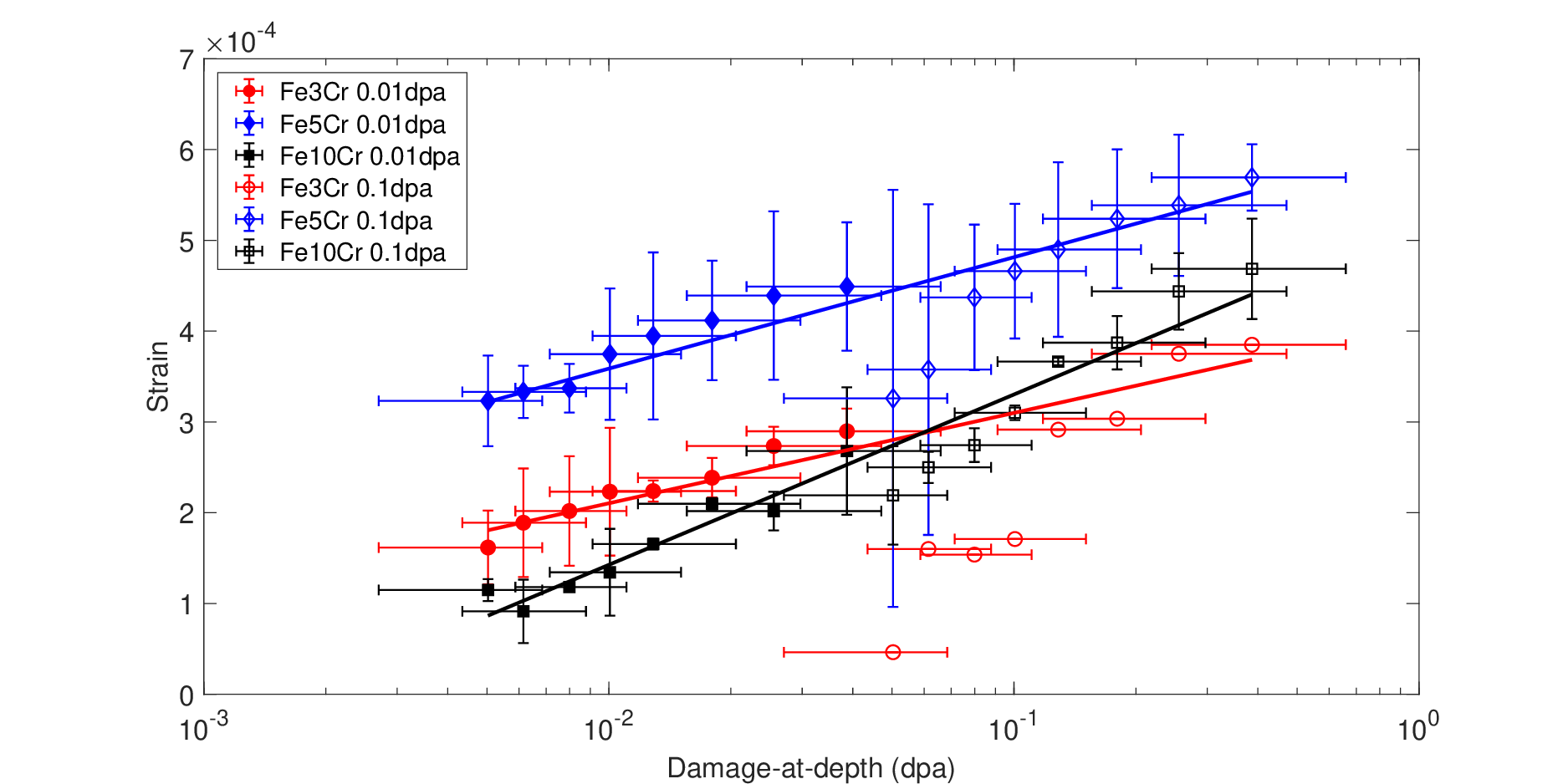}
\centering
\caption{The average measured strain as a function of damage at depth with the vertical error bars as $\pm$ one standard deviation of all the measurements on each sample. The damage-at-depth values are converted from the depth of the measurements using the damage profile calculated using SRIM. The horizontal error bars are calculated from the depth uncertainty in the depth measurements being $\pm 0.5 \mu$m.} 
\label{fig:Strain_Dpa}
\end{figure}

At each measured depth below the sample surface, the SRIM calculations shown in Figure \ref{fig:SRIM} provide us with a corresponding dpa value (the damage-at-depth). Using these depth-dependent dpa values, the strain (shown in Figure \ref{fig:Strain_Depth}) can be plotted as a function of the damage-at depth. The resulting plot is shown in Figure \ref{fig:Strain_Dpa} for all damage-at-depth levels considered. Fe5Cr exhibits the highest strain compared to Fe3Cr and Fe10Cr which show similar strain values. The mechanism behind this is unclear. It has been observed in several studies of neutron and self-ion irradiated FeCr, that for Cr content $< 20$\%, the greatest amount of macroscopic irradiation void swelling have been found in Fe9Cr \cite{Gelles1982, Katoh1995, Bhattacharya2016} and Fe6Cr \cite{Gelles1995, Sencer2000}. We have found lattice swelling to be greatest at Fe5Cr in this study, which is close to the findings in literature for macroscopic swelling. However, the effect of Cr on lattice strain and swelling still requires further investigation. 

The increase in strain with damage shows a consistent trend for all compositions, despite the measurements of damage-at-depth from 0.005 to 0.04 dpa and 0.05 to 0.4 dpa being taken on separate samples for each composition. The trends observed are qualitatively similar to the elastic strains measured in ion-irradiated UO$_{2}$, cubic-ZrO$_{2}$, MgO and ZrC \cite{Debelle2018}, where the strains at low damage levels ($< 0.5$ dpa) could be fitted to the approximated form:

\begin{equation} 
\label{eqn:strain}
\mathrm{strain} = m*\mathrm{ln(dpa)} + c 
\end{equation}

where $m$ corresponds to the rate of strain increase with damage. The constant $c$ offsets the fitting strain values trending towards negative infinity at low dpa in this approximated logarithm form. The values of the fit are shown in Table \ref{tab:strain_fit}. The rate of strain increase, $m$, is higher for larger Cr content suggesting higher defect retention with increasing Cr content, which agrees with TEM findings \cite{Jenkins2009, Prokhodtseva2013, Hernandez-Mayoral2008}.

\begin{table}
\begin{center}
\begin{tabular}{ |c|c|c| } 
\hline
Cr (wt \%)	&	$m$ $(\times 10^{-5})$	&	$c$ $(\times 10^{-4})$	\\
\hline
3	&	4.3 $\pm$ 0.5	&	4.1 $\pm$ 0.3	\\
5	&	5.3 $\pm$ 0.5	&	6.0 $\pm$ 0.4	\\
10	&	8.1 $\pm$ 0.4	&	5.2 $\pm$ 0.3	\\
\hline
\end{tabular}
\end{center}
\caption{Parameters fitted from experimental data to Equation \ref{eqn:strain}.}
\label{tab:strain_fit}
\end{table}

From simulations of BCC Fe at $T=0$ K \cite{Derlet2020}, it has been observed at low damage levels ($\sim$ 0.01 dpa) that Frenkel pairs form, but there is little clustering, leading to a rapid increase in total defect number density. With increasing damage (up to 0.5 dpa), the interstitials begin to cluster into loops and the growth in the number density of defects slows down. Then at even higher levels of damage ($>$ 2.5 dpa), extended dislocation networks are formed, reducing the number of isolated interstitials. It is likely that our experimental observations span the first two stages of defect evolution. We observe a rapid increase in strain at lower damage $\sim$ 0.01 dpa then a slower strain increase around 0.1 dpa, approximated by the logarithmic increase in strain with damage. It is also likely that we are observing defect clustering, and hence slower strain increase, at a lower damage level than in simulations, as small dislocation loops are mobile at room temperature. 

All the strains measured from the samples in this study are positive, corresponding to lattice expansion, suggesting that some interstitial defects have been retained after irradiation. We can use the measured lattice strain to estimate the equivalent Frenkel pair density in the materials. By considering the relative relaxation volumes ($\Omega_{r}$) of defects in pure Fe \cite{Ma2019}, the measured strain can be used to calculate the defect density in the material, by using the following expression \cite{Hofmann2015}:
\begin{equation}
\epsilon_{zz} = \frac{1}{3}\frac{(1+\nu)}{(1-\nu)} \sum_{A} n^{(A)} \Omega_{r}^{(A)}
\end{equation}
where $\epsilon_{zz}$ is the measured out-of-plane strain, $\nu = 0.3$ is the Poisson ratio (for pure Fe \cite{Cardarelli2018}), $n^{(A)}$ and $\Omega_{r}^{(A)}$ are respectively the number density and relative relaxation volume for each type of defect $(A)$. 

In FeCr, both $\langle 111 \rangle$ and $\langle 100 \rangle$ dislocation loops are observed and their relative proportions depend on the Cr content \cite{Yao2008, Schaublin2017}. The relative relaxation volume of a single self-interstitial defect (1 atom) in a $\langle 111 \rangle$ configuration is $\Omega_{r}^{\langle 111 \rangle} = 1.65$ and that of a $\langle 100 \rangle$ interstitial defect is $\Omega_{r}^{\langle 100 \rangle} = 1.86$ \cite{Ma2019}. These values correspond to isolated defects, not defects in clusters. We assumed that there is no clustering of interstitials, which is reasonable as we are examining low levels of damage, where large dislocation networks have not yet been formed \cite{Derlet2020}. If we were to account for the decrease in relaxation volume, per point defect, due to the clustering of interstitials \cite{Mason2019}, our estimated equivalent point defect density would increase. 

We will consider both the case of all interstitials being of $\langle 111 \rangle$ nature, and the case of all $\langle 100 \rangle$ interstitials. This will give the range of interstitial densities in the present FeCr samples. Furthermore, we need to consider the lattice strain due to the presence of vacancy defects, which have a relative relaxation volume of $\Omega_{r}^{v} = -0.22$ \cite{Ma2019}. At room temperature, vacancies are almost immobile and do not cluster \cite{Soneda1998}. The relaxation volume of a Frenkel pair can be taken as the sum of an interstitial and a vacancy \cite{Mason2019}. We also assume that no interstitials have been lost from the sample and that the number of interstitial atoms to be the same as the number of vacancies ($n^{i}$ $= n^{v}$), which is the number of equivalent Frenkel pairs. 

From the measured strains at 0.4 dpa of damage-at-depth, the volumetric number density of equivalent Frenkel pairs is $3.2-5.4\times 10^{25}$ m$^{-3}$. The lowest density is observed in Fe3Cr, assuming all interstitial defects are of $\langle 100 \rangle$ in nature, and the highest is in Fe5Cr, assuming all interstitial defects are of $\langle 111 \rangle$ in nature. For 0.04 dpa of damage-at-depth, the corresponding volumetric number density of equivalent Frenkel pairs is between $2.3-4.3\times 10^{25}$ m$^{-3}$. Note that these estimates provide a lower bound on the number of equivalent Frenkel pairs present in the irradiated material. We have estimated the total number of interstitial atoms (and vacancies) in the damaged material, whether they exist as isolated interstitials or in clusters. We did not account for the reduction in relaxation volume per point defect ($\Omega_{r}$) if interstitials existed in clusters. In the case of defect clustering, to produce the same strain from dislocation networks as isolated defects, a higher equivalent Frenkel pair number density would be required. Similarly, in assuming no loss of interstitials, which provide the positive contributions to the lattice strain, we have estimated the minimum number of equivalent Frenkel pairs required to produce the measured strain. Further details of this calculation, and of the TEM comparisons to follow, can be found in the supplementary file.


Previous TEM studies of FeCr found an areal defect density of around $1\times 10^{15}$ m$^{-2}$ at 0.3 dpa \cite{Yao2008}. The average diameter of the loops observed was 2 nm from measurements made in weak-beam dark field conditions with \textbf{g} = $\{110\}$. The thickness of the damaged layer was 19 nm, from irradiation with 100 keV Fe$^{+}$ ions, as determined by SRIM calculations \cite{Yao2008}. We assume that all imaged loops are of interstitial nature. This can be justified as vacancies have low mobility and do not cluster significantly at room temperature. In making this assumption, the following estimates provide an upper bound on the equivalent Frenkel pair density observed in TEM. 

Consider for BCC Fe, the lattice parameter is $a_{0}$ = 0.287 nm, the $(111)$ and $(100)$ planar densities are respectively $\sim$7 atoms nm$^{-2}$ and $\sim$12 atoms nm$^{-2}$. Again, we consider the two extreme cases of all $\langle 111 \rangle$ loops and all $\langle 100 \rangle$ loops to obtain a range of the possible interstitial densities. We also account for the loops that are invisible when $\mathbf{g\cdot b} = 0$, which includes $\tfrac{1}{2}$ of the $\langle 111 \rangle$ loops and $\tfrac{1}{3}$ of the $\langle 100 \rangle$ loops. This gives the volumetric number density of interstitial atoms, and hence equivalent Frenkel pairs, to be between $2.3 - 3.0 \times 10^{24}$ m$^{-3}$. This value is an order of magnitude lower than the densities obtained from our lattice strain measurements.

At lower doses, we can compare with TEM measurements from a study with damage level of 0.05 dpa on FeCr which found defect densities of up to $1 \times 10^{23}$ m$^{-3}$ \cite{Schaublin2017}. The authors of that study have already corrected for the cases of $\mathbf{g\cdot b} = 0$. In the irradiated Fe5Cr and Fe10Cr (which are the same materials as this study), the reported average defect size is 1.1 nm. Using similar calculations as above, this gives the number density of interstitial atoms, and hence equivalent Frenkel pairs, to be $6.7 - 11.4 \times 10^{23}$ m$^{-3}$. This point defect density obtained from TEM measurements is over an order of magnitude lower than the values estimated from our lattice strain measurements.


The discrepancy in defect densities between TEM and lattice strain measurements may be partially accounted for by the loss of defect loops to the surface of TEM samples. Another factor to consider is contributions from defects below the sensitivity limit of TEM \cite{Phillips2020}. One point that supports this is that in TEM measurements, the density of defects increases as $(dpa)^{n}$ with $1 \leq n \leq 2$ \cite{Yao2008}, whereas the defect density calculated from strain measurements is proportional to the natural logarithm of damage. This suggests that TEM measurements could be underestimating the defect density particularly at lower damage levels where the defects are smaller. \textit{Ab initio} calculations have shown that for small clusters of fewer than 51 self-interstitial atoms in Fe, C15-Laves phase clusters \cite{Marinica2012} are the most stable configuration but their size ($\sim$ 1.5 nm diameter) is not accessible to TEM \cite{Alexander2016}. The presence of these clusters would still impart measurable strain to the material \cite{Zhang2015} contributing to the discrepancy between the measured strain from this work and the defect density measured in TEM. The measurement of lattice strain from damage levels below 0.01 dpa shows that the `threshold damage' derived from TEM studies \cite{Yao2008, Jenkins1978, Kirk1987} likely results from the lack of visibility of small defects in TEM rather than an inherent `threshold' for damage formation. This is also supported by more recent research \cite{Schaublin2017} which revealed that the dislocation loops were visible in TEM at 0.0015 dpa for Fe5Cr, Fe10Cr (both are the same material as used in this study), Fe14Cr and pure Fe. Correlation between lattice strain and defect contrast in TEM offer a promising way forward for fully describing the irradiation-induced defect population at low damage levels.

\subsection*{Comparing Trends}
The measurements of hardness, thermal diffusivity and lattice strain in this study were conducted on the same set of samples, however different material properties resulting from the same damage microstructure exhibit different sensitivity to Cr content and irradiation dose. All three material properties show that the amount of Cr in the material has a significant effect on the response of the material to irradiation. Fe5Cr and Fe10Cr samples exhibit more change in hardening, thermal diffusivity and rate of lattice strain increase after irradiation than their Fe3Cr counterpart. This is attributed to the role of Cr in reducing defect mobility and thus enhancing total defect retention in the material, which has previously been observed in TEM measurements \cite{Prokhodtseva2013, Yao2008}. 

The absolute values of hardness and thermal diffusivity at each damage level vary monotonically with Cr content, however, the absolute value of lattice strain is the greatest for Fe5Cr regardless of damage level. This non-monotonic variation in the lattice strain with Cr content could be due to the influence of Cr on the retention of each type of defect (interstitial vs. vacancy). Lattice strain is the result of the net contributions of position relaxation volume from interstitial defects and negative relaxation volume from vacancies. As such, it is sensitive to the population of different defect types, as well as extended defect structures \cite{Das2020}, rather than the total defect population \cite{Derlet2020}.

Irradiation hardening and lattice strain are both very sensitive to the presence of irradiation damage in the material. The effect of damage as low as 0.01 dpa can clearly be quantified in both these properties. The changes are very significant after 0.1 dpa of nominal damage, with Fe5Cr and Fe10Cr exhibiting over 46\% increase in hardness and lattice strain of up to 5$\times$10$^{-4}$. This shows that the presence of the irradiation-induced defects plays a large role in the changes of these properties. Since defect retention is increased by the presence of Cr atoms, which in itself also affects hardness and lattice strain, these material properties are hence sensitive to both Cr content and damage levels.

In contrast, the decrease in thermal diffusivity of FeCr can only be measured after 0.1 dpa of damage, and even then, the changes are quite small ($\sim$ 8\% for Fe5Cr and Fe10Cr). The change to thermal diffusivity is dominated by the Cr content in the material, as the Cr atoms, as well as the defects, act as electron scattering sites. The lowest order scattering rate, in the dilute limit, of a Frenkel pair in Fe is 19 fs$^{-1}$ (calculations included in Appendix E), which is much higher than that of a Cr atom at 2.9 fs$^{-1}$. However, since Cr atoms are also present in much higher concentration than the defects, this likely explains the greater role of Cr content than damage level on thermal diffusivity degradation. 


\section*{Conclusion}
We have studied Fe3Cr, Fe5Cr and Fe10Cr implanted with 20 MeV Fe$^{3+}$ ions to nominal damage levels of 0.01 dpa and 0.1 dpa at room temperature. The hardness, thermal conductivity and lattice strain were probed with nanoindentation, transient grating spectroscopy and X-ray micro-beam Laue diffraction respectively. From these findings, we conclude the following:
\begin{itemize}
	\item All irradiated samples exhibited hardening with measurable changes observed at nominal damage level as low as 0.01 dpa. 
	\item Fe5Cr and Fe10Cr samples show greater irradiation hardening at each nominal damage level than Fe3Cr. This is attributed to higher defect retention induced by increasing Cr content. The shape of the hardness vs. indentation depth curves qualitatively agrees with an existing plastic zone model. 
	\item Thermal diffusivity of the materials could be successfully measured by TGS. The values from unirradiated Fe3Cr and Fe5Cr agree with a kinetic theory model that attributes the degradation in thermal diffusivity to Cr atoms acting as electron scattering sites. 
	\item The thermal diffusivity of Fe5Cr and Fe10Cr showed an 8\% reduction after 0.1 dpa of nominal damage, while Fe3Cr samples exhibited negligible change after irradiation. After irradiation to 0.01 dpa, Fe3Cr and Fe10Cr showed small increases in thermal diffusivity. The reduction of thermal diffusivity after 0.1 dpa is probably due to Cr atoms causing a greater retention of irradiation-induced defects, which also act as electron scattering sites. Overall the changes in thermal diffusivity as a function of irradiation damage are small.
	\item Measurable strain was observed for damage-at-depth levels below 0.01 dpa, and lattice strain of over $5 \times 10^{-4}$ is measured in FeCr at a damage level of 0.4 dpa in damage-at-depth. Although lattice strain showed different dependence on Cr content between irradiations to 0.01 and 0.1 dpa, the rate of increase in lattice strain is higher for greater Cr content. This is again attributed to greater defect retention due to Cr atoms.
	\item The lower bound of defect density in the irradiated FeCr, calculated from the measured strain, is higher than that previously observed in TEM studies of corresponding damage levels. This indicates that both lattice strain and TEM measurements are crucial to fully characterise the  irradiation-induced defect population at low doses.
\end{itemize}

All data, raw and processed, as well as the processing and plotting scripts used in this study are available online at: \url{https://doi.org/10.5287/bodleian:xvNyeYP0z}

\section*{Acknowledgements}
The authors would like to thank Sergei Dudarev and Andrew London from the Culham Centre for Fusion Energy for their helpful discussions. The authors would also like to thank Jon Tischler and Wenjun Liu for their discussions and support at the 34-ID-E beamline at the Advance Photon Source, Bo-Shiuan Li and Naganand Saravanan from Oxford Materials Department for their assistance with nanoindentation and EBSD measurements. This work was funded by Leverhulme Trust Research Project Grant RPG-2016-190. X-ray diffraction measurements were performed at the Advanced Photon Source, a U.S. Department of Energy (DOE) Office of Science User Facility operated for the DOE Office of Science by Argonne National Laboratory under Contract No. DE-AC02-06CH11357. KS acknowledges funding from the General Sir John Monash Foundation and Oxford Engineering Science Department.  NWP and FH acknowledge funding from the European Research Council (ERC) under the European Union's Horizon 2020 research and innovation programme (grant agreement No. 714697). DEJA acknowledges funding from EPSRC grant EP/P001645/1.

\newpage
\appendix

\renewcommand{\thefigure}{B-\arabic{figure}}

\setcounter{figure}{0}
\section*{Appendix A - Material Composition}
Chemical analysis of the alloys in the as-delivered final metallurgical condition \cite{Coze2007}.

\begin{center}
\begin{tabular}{ |c|c|c|c|c|c|c| } 
\hline
Alloy	&	C wt (ppm)	&	S wt (ppm)	&	O wt (ppm)	&	N wt (ppm)	&	P wt (ppm)	&	Cr wt\%	\\
\hline
Fe3Cr	&	4	&	3	&	6	&	2	&	-	&	3.05	\\
Fe5Cr	&	4	&	3	&	6	&	2	&	$<$ 5	&	5.40	\\
Fe10Cr	&	4	&	4	&	4	&	3	&	$<$ 5	&	10.10	\\

\hline
\end{tabular}
\end{center}

\section*{Appendix B - Elastic Modulus Data}
\begin{figure}[h!]
	\includegraphics[width=\textwidth]{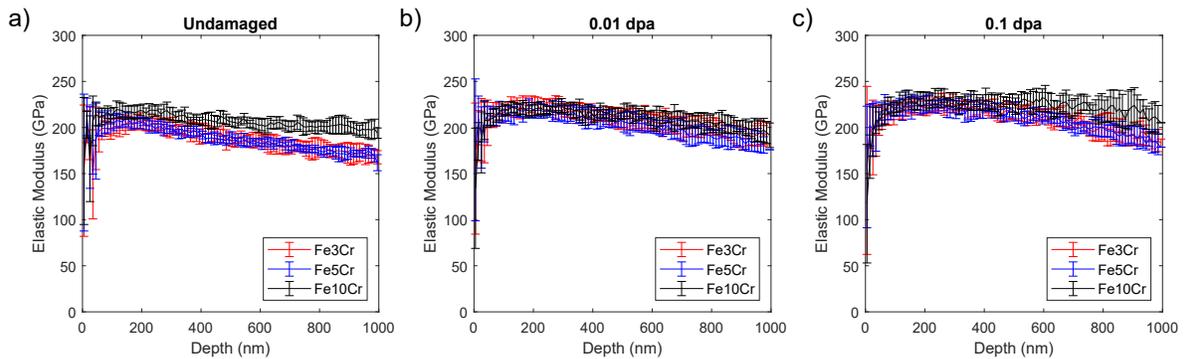}
	\centering
	\caption{Elastic modulus of the samples as a function of indented depth from CSM nanoindentation measurements. The data is grouped by damage level allowing comparison of the elastic modulus across different Cr content.} 
	\label{fig:Modulus_Depth}
\end{figure}

The data shown in Figure \ref{fig:Modulus_Depth} is the average elastic from all indents taken on each sample, with the error bars representing $\pm$ one standard deviation. It can be seen that for the undamaged samples and samples with 0.1 dpa of nominal damage, the elastic modulus for Fe3Cr and Fe5Cr do not show a difference at any indentation depth. Fe10Cr exhibit slightly higher values at depths greater than 400 nm. However, these relative difference are small compared to the difference in hardness between samples of different Cr content (Figure \ref{fig:Hardness_Depth}a). For 0.01 dpa of nominal damage, the elastic modulus of all samples do not show any significant difference, within experimental errors.

\renewcommand{\theequation}{C-\arabic{equation}}
\setcounter{equation}{0}  
\section*{Appendix C - Experimental Validation of the Johnson Model}
The Johnson model from Equation \ref{eqn:pz}, modified for a Berkovich tip, is:
\begin{equation}
	\frac{z_{p}}{h} = \left(\frac{2E}{3\sigma_{ys}}\right)^{1/3}
\end{equation}
where $z_{p}$ is the depth of the plastic zone directly beneath the indenter, $h$ is the indentation depth, $E$ is Young's modulus and $\sigma_{ys}$ is the yield strength. The ratio $\frac{z_{p}}{h}$ can be obtained by looking at $h_{c}$, which is the indentation depth at which the plastic zone penetrates into the unirradiated bulk (which is known to be 3.5 $\mu$m below the surface from SRIM calculations). Therefore $\frac{z_{p}}{h} = \frac{3.5}{h_{c}}$, where $h_{c}$ is in $\mu$m.

The yield strength, $\sigma_{ys}$, was obtained from a study by Matijasevic \textit{et al.} \cite{Matijasevic2008} with similar Cr content (2.5\%, 5\% and 9\%Cr). The results are below:
\begin{table}[h!]
\begin{center}
	\begin{tabular}{ |P{1.5cm}|P{3cm}|P{1cm}|P{2.5cm}|P{3.5cm}|P{1.5cm}| } 
		\hline
		Alloy	&	$h_{c}$ 	&	$\frac{z_{p}}{h}$	&	$\sigma_{ys}$ (MPa)	&	$E$ (GPa)	&	$\left(\frac{2E}{3\sigma_{ys}}\right)^{1/3}$	\\
		& (from 0.1 dpa samples) & 				& 	(from \cite{Matijasevic2008})				& (averaged between 300-600 nm) & \\
		\hline
		Fe3Cr	&	565	&	4.31	&	409 (Fe2.5Cr)	&	228	&	7.19	\\
		Fe5Cr	&	525	&	4.71	&	462	(Fe5Cr) &	218	&	6.80	\\
		Fe10Cr	&	745	&	3.03	&	514	(Fe9Cr) &	218	&	6.56	\\
			
		\hline
	\end{tabular}
\end{center}
\end{table}

We can conclude that fitting our experimental data to the Johnson model yields sensible results that agree within a factor of 2 to 3.

\section*{Appendix D - Parameters for Kinetic Theory Model}
The parameters used to fit the kinetic theory model, in Equation \ref{eqn: kinetic}, to the experimental data for thermal diffusivity.
\begin{center}
\begin{tabular}{ |c|c|c|c| } 
\hline
Symbol	&	Quantity	&	Value	&	Source	\\
\hline
$C_{e}/T$	&	Electronic Heat Capacity Coefficient of Fe	&	$2.095 \times 10^{-9}$ eV K$^{-2}$ \AA{}$^{-3}$	&	\cite{Mason2015}	\\
$\nu_{f}$	&	Fermi Velocity of Fe	&	$4.95$ \AA{} fs$^{-1}$	&	\cite{Mason2015}	\\
$c_{p}$	&	Specific Heat Capacity of Fe	&	$25.14$ J mol$^{-1}$ K$^{-1}$	&	\cite{Chase1998}	\\
$\rho$	&	Density of Fe	&	$7850$ kg m$^{-3}$	&	\cite{Chase1998}	\\
$\kappa_{Fe}$	&	Thermal Conductivity of Fe	&	$75.8$ W m$^{-1}$ K$^{-1}$	&	\cite{Fulkerson1966}	\\

\hline
\end{tabular}
\end{center}

\renewcommand{\theequation}{E-\arabic{equation}}
\setcounter{equation}{0}  
\section*{Appendix E - Calculation of Frenkel Pair Scattering Rate in Fe}
In a similar form to Equation \ref{eqn: kinetic}, the thermal conductivity, $\kappa$, due to scattering from Frenkel pairs in pure Fe can be written as:
\begin{equation}
	\kappa = \frac{1}{3} C_{e}\nu_{f}{}^{2} \left( \frac{c_{FP}}{\tau_{e,FP}} + \frac{1-c_{FP}}{\tau_{e,Fe}} \right)^{-1}
\end{equation} \
where $C_{e}$ is the electronic heat capacity, $\nu_{f}$ is the Fermi velocity, $c_{FP}$ is the Frenkel pair concentration, $\tau_{e,FP}$ is the electron scattering time from the Frenkel pairs and $\tau_{e,Fe}$ is the scattering time of pure Fe.

Using the Wiedemann-Franz law, the above equation can be written as:
\begin{equation}
	\rho_{e} = \frac{3LT}{C_{e}\nu_{f}{}^{2} } \left( \frac{c_{FP}}{\tau_{e,FP}} + \frac{1-c_{FP}}{\tau_{e,Fe}} \right)
\end{equation}
where $\rho_{e}$ is the electrical resistivity, $L$ is the Lorenz number and $T$ is the temperature. Differentiating both sides by $c_{FP}$, we obtain an expression involving resistivity per Frenkel pair. This has an experimental value of 20 $\mu\Omega$m \cite{Dimitrov1984}. This allows us to solve for $\tau_{e, FP}$, which has a value of 0.052 fs.

\newpage

\renewcommand{\thefigure}{S-\arabic{figure}}
\setcounter{figure}{0}
\renewcommand{\theequation}{S-\arabic{equation}}
\setcounter{equation}{0}
\section*{Supplementary File - Defect Density Calculations and Comparisons}
\subsection*{1. Defect Density From Lattice Strain Measurements}

We use the following expression \cite{Hofmann2015} for strain as a function of defect density:
\begin{equation}\label{eqn:supstrain}
\epsilon_{zz} = \frac{1}{3}\frac{(1+\nu)}{(1-\nu)} \sum_{A} n^{(A)} \Omega_{r}^{(A)}
\end{equation}
where $\epsilon_{zz}$ is the measured out-of-plane strain, $\nu = 0.3$ is the Poisson ratio (for pure Fe \cite{Cardarelli2018}), $n^{(A)}$ and $\Omega_{r}^{(A)}$ are respectively the number density and relative relaxation volume for each type of defect $(A)$. 

As the lattice strain values measured are all positive, this means some interstitial defects are present in our material. We then make the following assumptions in our calculations:
\begin{itemize}
	\item There is no clustering of interstitials. This means the relaxation volume per point defect is maximised \cite{Mason2019}.
	\item There is no loss of interstitial defects, which are much more mobile than vacancies \cite{Soneda1998}, to the surface of the materials. This means there are equal numbers of interstitial atoms ($n^{i}$) and vacancies ($n^{v}$).
\end{itemize}
By employing these assumptions, we obtain the lower bound of the equivalent Frenkel pair density ($n^{FP}$). If interstitials did cluster, then the relaxation volume per point defect would decrease, requiring more Frenkel pairs to be present in order to produce the same amount of positive strain that was measured (covered by assumption 1). If interstitials were lost to the surface, again more equivalent Frenkel pairs would need to be present to account for the amount of positive strain measured (covered by assumption 2).

Rearranging Equation \ref{eqn:supstrain}, we get:
\begin{equation}\label{eqn:density}
n^{FP} = \frac{\epsilon_{zz}}{\Omega_{r}^{FP}} \left(\frac{3(1-\nu)}{(1+\nu)}\right)
\end{equation}

The relative relaxation volume, per point defect, of a $\langle 111 \rangle$ interstitial is $\Omega_{r}^{\langle 111 \rangle} = 1.65$ and that of a $\langle 100 \rangle$ interstitial defect is $\Omega_{r}^{\langle 100 \rangle} = 1.86$ \cite{Ma2019}. For a vacancy, the relative relaxation volume is $\Omega_{r}^{v} = -0.22$ \cite{Ma2019}. In order to obtain the volumetric number density ($N^{FP}$) of the equivalent Frenkel pairs, we then need to multiply by the atomic density of Fe ($\rho_{Fe} = 8.48 \times 10^{28}$ m$^{-3}$).

\begin{figure}[h!]
	\includegraphics[width=0.8\textwidth]{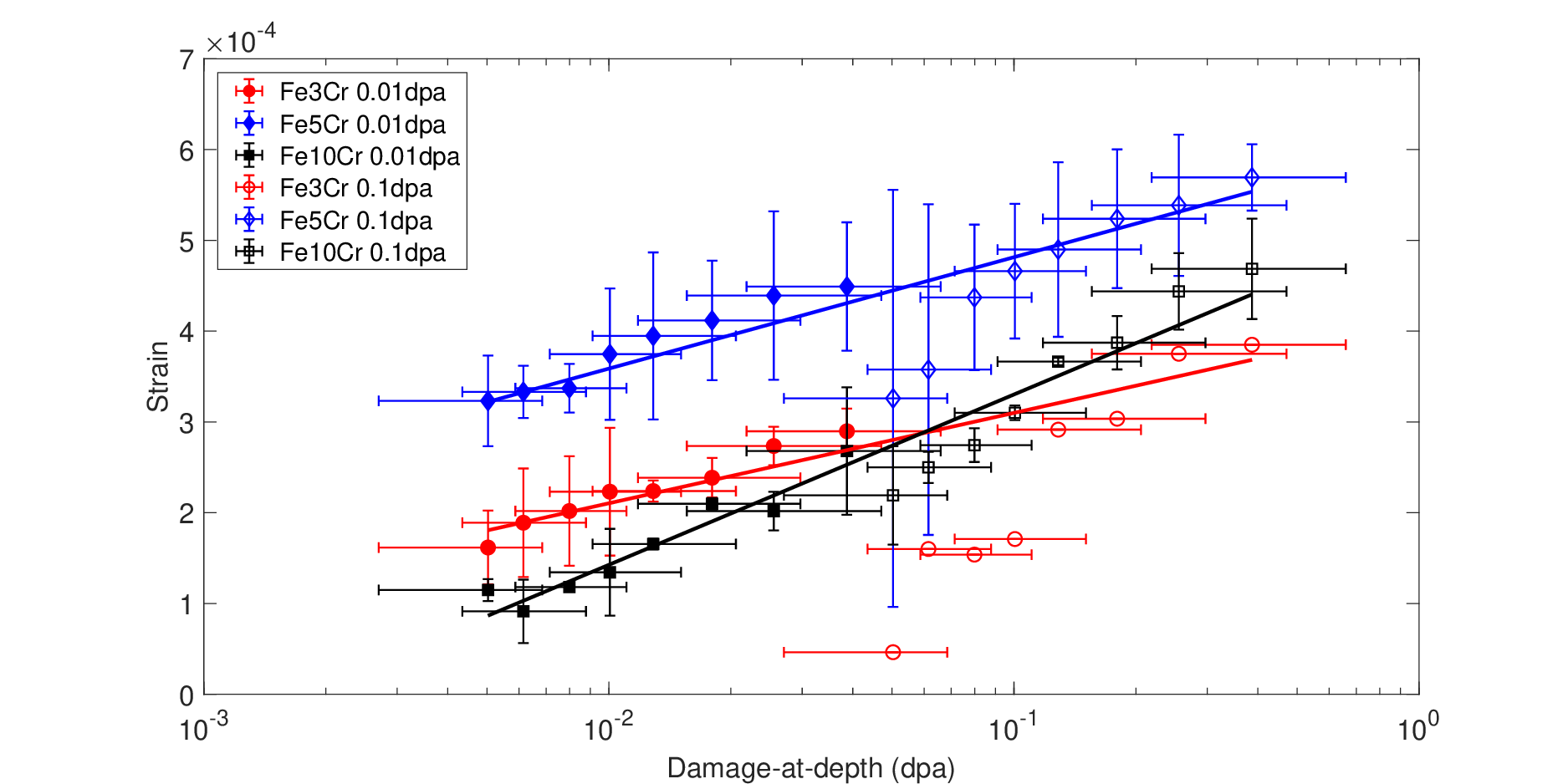}
	\centering
	\caption{Data from Figure 7 of our work.}  
	\label{fig:strain}
\end{figure}

\underline{An example calculation:}
At 0.04 dpa of damage-at-depth, the largest strain observed is from Fe5Cr at $4.491 \times 10^{-4}$ (from Figure \ref{fig:strain}). If all the interstitial defects are of $\langle 111 \rangle$ nature, then $\Omega_{r}^{FP, \langle 111 \rangle} = 1.65 - 0.22 = 1.43$. Substituting this into Equation \ref{eqn:density}:
\begin{align*}
N^{FP, \langle 111 \rangle, Fe5Cr} &= \rho_{Fe} \times n^{FP, \langle 111 \rangle, Fe5Cr}  \\
&= \rho_{Fe} \times \frac{\epsilon_{zz}^{Fe5Cr}}{\Omega_{r}^{FP, \langle 111 \rangle}} \times \left(\frac{3(1-\nu)}{(1+\nu)}\right) \\
&= (8.48 \times 10^{28}) \times \left(\frac{4.49 \times 10^{-4}}{1.43}\right) \times \left(\frac{3(1-0.3)}{(1+0.3)}\right) \\
&= 4.29 \times 10^{25} \text{ m}^{-3}
\end{align*}

As Fe5Cr has the highest strain, and $\langle 111 \rangle$ defects have the lowest relaxation volume per defect, this value is the upper range of our equivalent Frenkel pair density estimate at 0.04 dpa of damage-at-depth. For the lower range, this will come from the calculation using the strain value of Fe3Cr ($\epsilon_{zz}^{Fe3Cr} = 2.69 \times 10^{-4}$ from Figure \ref{fig:strain}) and assuming all interstitial defects are of $\langle 100 \rangle$ nature ($\Omega_{r}^{FP, \langle 100 \rangle} = 1.86 - 0.22 = 1.64$). Using the same calculation steps as above, we obtain $N^{FP, \langle 100 \rangle, Fe3Cr} = 2.25 \times 10^{25}$ m$^{-3}$. 

For 0.4 dpa of damage-at-depth, we repeat the same calculations, using $\epsilon_{zz}^{Fe5Cr} = 5.69 \times 10^{-4}$ and $\epsilon_{zz}^{Fe3Cr} = 3.85 \times 10^{-4}$  (from Figure \ref{fig:strain}). We get the range of equivalent Frenkel pair volumetric number density $3.22 - 5.44 \times 10^{25}$ m$^{-3}$.

To summarise, from our lattice strain measurements, we estimate equivalent Frenkel pair volumetric number densities:
\begin{itemize}
	\item At 0.04 dpa of damage-at-depth: $2.25 - 4.29 \times 10^{25}$ m$^{-3}$
	\item At 0.4 dpa of damage-at-depth: $3.22 - 5.44 \times 10^{25}$ m$^{-3}$
\end{itemize}

\subsection*{2. Defect Density from TEM Studies}
For calculations of equivalent Frenkel pair density from TEM studies, we make the following assumption:
\begin{itemize}
	\item All dislocation loops and defects observed are of interstitial nature.
\end{itemize}
TEM studies have found that a majority of the defects observed are interstitials rather than vacancies \cite{Xu2009, Schaublin2017}. This is generally attributed to the lower mobility of vacancies, leading to less growth of vacancy loops to sizes observable in TEM. By assuming all observable defects are of interstitial nature, we calculate the upper bound of the equivalent Frenkel pair density in the material.

\subsubsection*{2.1. From Yao \textit{et al.} - 0.4 dpa}
\begin{figure}[h!]
	\includegraphics[width=0.8\textwidth]{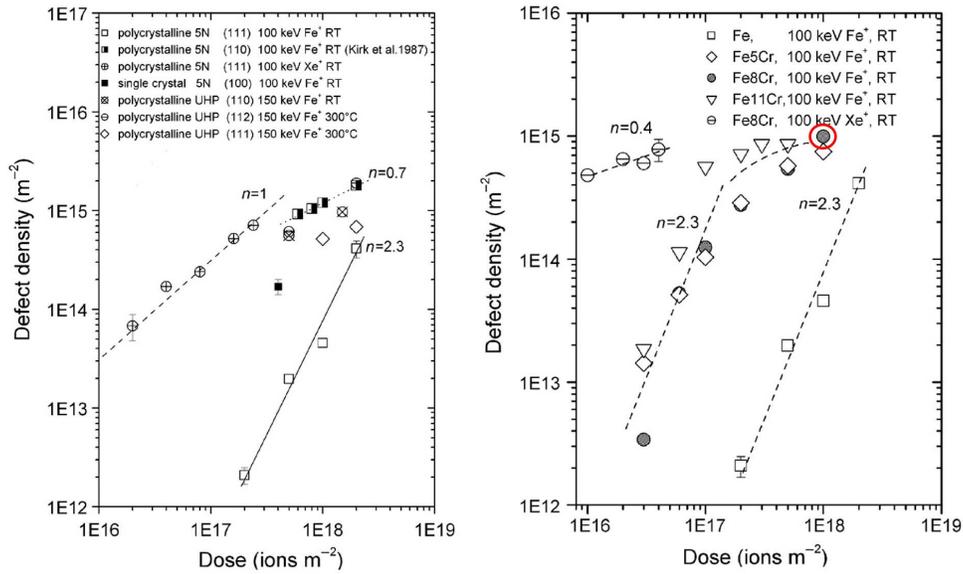}
	\centering
	\caption{Data from Figure 2 of \cite{Yao2008}. Yao \textit{et al.} quoted $2 \times 10^{16}$ ions m$^{-2}$ as being equivalent to 0.01 dpa. The defect density for FeCr at $8 \times 10^{17}$ ions m$^{-2}$ is then taken to be approximately $1 \times 10^{15}$ m$^{-2}$. The red circle was added to indicate the data point from which the value was taken.}  
	\label{fig:yao}
\end{figure}

\begin{figure}[h!]
	\includegraphics[width=0.5\textwidth]{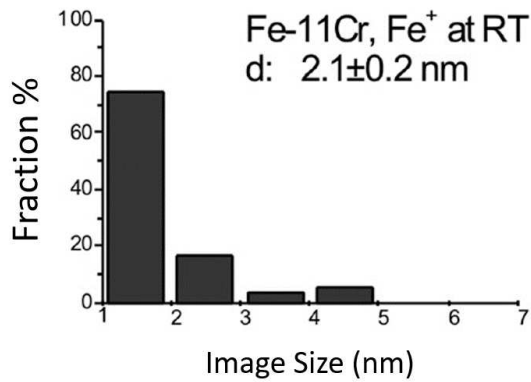}
	\centering
	\caption{Data from Figure 6 of \cite{Yao2008}. The only reported values for defect sizes in FeCr were of Fe11Cr. The damage level is equivalent to 0.15 dpa but the authors reported similar distribution of sizes for damage levels below 0.5 dpa.}  
	\label{fig:yao2}
\end{figure}

Yao \textit{et al.} \cite{Yao2008} found the following:
\begin{itemize}
	\item Areal defect density = $1 \times 10^{15}$ m$^{-2}$ (see Figure \ref{fig:yao}).
	\item Average defect diameter = 2.1 nm (see Figure \ref{fig:yao2}).
	\item Damage layer thickness = 19 nm.
\end{itemize}
For simplicity, only one value of damage level was compared. 0.4 dpa was chosen as it has the largest difference in damage compared to the other TEM study in reference \cite{Schaublin2017}.

The lattice parameter of BCC Fe is $a_{0} = 0.287$ nm. The planar density of the (100) plane is $\frac{1}{a_{0}{}^{2}} = 12.1$ atoms nm$^{-2}$. For the (111) plane, the planar density is $\frac{1}{\sqrt{3}a_{0}{}^{2}} = 7.0$ atoms nm$^{-2}$. We also need to take into account the defects that are invisible due to $\mathbf{g}\cdot \mathbf{b} = 0$. In the TEM study, $\mathbf{g} = \{110\}$, which means only $\tfrac{1}{2}$ of the $\langle 111 \rangle$ defects and $\frac{2}{3}$ of the $\langle 100 \rangle$ defects are actually visible.

The equivalent Frenkel pair volumetric density ($N^{FP}$) can be calculated by:
\begin{equation}
N^{FP} = \frac{\text{areal defect density} \times \text{defect area} \times \text{planar density}}{\text{damage layer thickness}\times \text{fraction of defects visible}}
\end{equation}

If all interstitials were of $\langle 111 \rangle$ nature:
\begin{align*}
N^{FP, \langle 111 \rangle} &= \frac{(1 \times 10^{15}) \times ((1.05\times 10^{-9})^{2}\times \pi) \times (7.0\times 10^{18})}{19\times 10^{-9}\times \frac{1}{2}} \\
&= 2.6 \times 10^{24} \text{ m}^{-3}
\end{align*}

Similarly, if all interstitials were of $\langle 100 \rangle$ nature:
\begin{align*}
N^{FP, \langle 100 \rangle} &= \frac{(1 \times 10^{15}) \times ((1.05\times 10^{-9})^{2}\times \pi) \times (12.1\times 10^{18})}{19\times 10^{-9}\times \frac{2}{3}} \\
&= 3.3 \times 10^{24} \text{ m}^{-3}
\end{align*}

\subsubsection*{2.2. From Sch\"{a}ublin \textit{et al.} - 0.05 dpa}
\begin{figure}[h!]
	\includegraphics[width=0.6\textwidth]{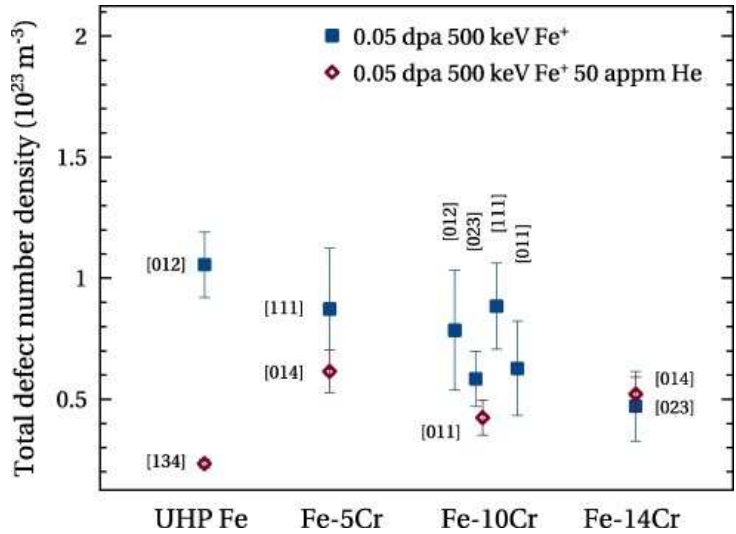}
	\centering
	\caption{Data from Figure 5 of \cite{Schaublin2017}. At 0.05 dpa (Fe$^{+}$ irradiation only), the highest defect density observed in FeCr is $8 \times 10^{23}$ m$^{-3}$.}  
	\label{fig:schaublin}
\end{figure}

\begin{figure}[h!]
	\includegraphics[width=0.8\textwidth]{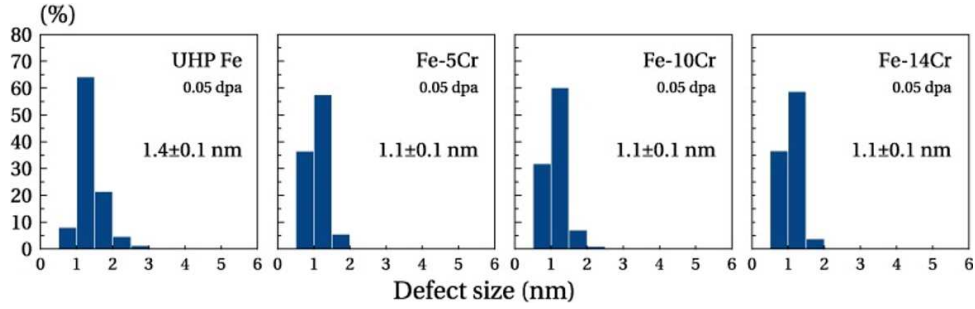}
	\centering
	\caption{Data from Figure 6 of \cite{Schaublin2017}. At 0.05 dpa (Fe$^{+}$ irradiation only), the average defect size in FeCr is 1.1 nm.}  
	\label{fig:schaublin2}
\end{figure}

Sch\"{a}ublin \textit{et al.} \cite{Schaublin2017} found the following:
\begin{itemize}
	\item Volume defect density = $1 \times 10^{23}$ m$^{-3}$ (see Figure \ref{fig:schaublin}, corrections for $\mathbf{g}\cdot \mathbf{b} = 0$ already made).
	\item Average defect diameter = 1.1 nm (see Figure \ref{fig:schaublin2}).
\end{itemize}

The equivalent Frenkel pair volumetric density ($n_{d}$) can be calculated by:
\begin{equation}
N^{FP} = \text{volume defect density} \times \text{defect area} \times \text{planar density}
\end{equation}

If all interstitials were of $\langle 111 \rangle$ nature:
\begin{align*}
N^{FP, \langle 111 \rangle} &= (1 \times 10^{23}) \times ((0.55\times 10^{-9})^{2}\times \pi) \times (7.0\times 10^{18}) \\
&= 6.65 \times 10^{23} \text{ m}^{-3}
\end{align*}

Similarly, if all interstitials were of $\langle 100 \rangle$ nature:
\begin{align*}
N^{FP, \langle 100 \rangle} &=(1 \times 10^{23}) \times ((0.55\times 10^{-9})^{2}\times \pi) \times (12.1\times 10^{18}) \\
&= 1.15 \times 10^{24} \text{ m}^{-3}
\end{align*}

\subsection*{3. Comparisons}
The following table summarises the equivalent Frenkel pair densities from lattice strain measurements and TEM studies:

\begin{center}
	\begin{tabular}{ |c|c|c| } 
		\hline
		Damage (dpa)	&	Lattice Strain Measurements ($\times 10^{23}$ m$^{-3}$)	&	TEM Measurements ($\times 10^{23}$ m$^{-3}$)	\\
		\hline
		0.04 - 0.05	&	225	- 429 &	6.7 - 11.5	\\
		0.4	&	322 - 544	&	26 - 33	\\
		\hline
	\end{tabular}
\end{center}

\newpage
\bibliographystyle{elsarticle-num}
\bibliography{ref}

\end{document}